\documentclass[11pt]{amsart}

\usepackage{qzart}
\usepackage{multirow,booktabs,array}

\tikzstyle{block} = [rectangle, draw, text width=6em, text centered,
rounded corners, minimum height=3em]
\tikzstyle{line} = [draw, -latex']

\begin{document}

\title[Sensitivity Value]{On Sensitivity Value of Pair-Matched
  Observational Studies}

\author{Qingyuan Zhao}
\address{Department of Statistics, University of
  Pennsylvania}

\email{qyzhao@wharton.upenn.edu}


\maketitle

\begin{abstract}
An observational study may be biased for estimating causal effects
by failing to control for unmeasured confounders. 
This paper proposes a new
quantity called the ``sensitivity value'', which is defined as the
minimum strength of unmeasured confounders needed to change
the qualitative conclusions of a naive analysis assuming no
unmeasured confounder. We establish the asymptotic normality of
the sensitivity value in pair-matched observational studies. The theoretical
results are then used to approximate the power of a sensitivity analysis and select
the design of a study. We explore the potential to use sensitivity values to
screen multiple hypotheses in presence of unmeasured confounding using
a microarray dataset.


\end{abstract}

\section{Introduction}
\label{sec:introduction}

In a pair-matched observational study, subjects are matched by their
observed covariates, but the difference within a matched pair could
still be due to unmeasured confounders instead of a genuine treatment
effect. To study how sensitive the qualitative conclusions (in this paper
significance of the treatment effect) are to unmeasured confounders, a
commonly used model of \citet[Chapter 4]{rosenbaum2002observational}
uses a single parameter $\Gamma$ to
represent the magnitude of departure from random assignment;
$\Gamma=1$ means random assignment and larger $\Gamma$ means a larger departure from random assignment.
In such sensitivity analyses, the user typically
computes the range of $p$-values
$[\underline{p}_{\Gamma},\overline{p}_{\Gamma}]$ under different
levels of $\Gamma$. When $\Gamma = 1$, $\underline{p}_{\Gamma} =
\overline{p}_{\Gamma}$ and they are equal to the usual $p$-value under the null hypothesis.

We illustrate the typical process of sensitivity analysis using a
microarray dataset. This microarray experiment investigates where
genes are differentially expressed in human brain with respect to
gender \citep{vawter2004}. By assuming a linear structural model
between gene expressions, gender and unmeasured confounders,
\citet{gagnon2012} and \citet{wang2016confounder}
studied the dataset and found evidence of serious unmeasured
confounding. The 84 observations in this dataset were obtained from
three different laboratories on two different microarray platforms. To
form a pair-matched observational study, we match the observations
exactly by the lab and platform and obtain 41 pairs of males and
females.

To assess which genes are differentially expressed in
males and females, we can use Wilcoxon's signed rank test to compute a
$p$-value for each of the $12,600$ genes in the dataset.
A sensitivity analysis augments the significance test by considering
possible departures from random assignment. \Cref{tab:gene} shows the
sensitivity analysis of $9$ probe sets in the dataset. The
$\Gamma = 1$ column corresponds to the usual Wilcoxon's signed rank
test. All the $5$ probe sets shown in \Cref{tab:gene} have very small
two-sided $p$-values ($<0.01$). As we increase the sensitivity parameter $\Gamma$, the $p$-value upper
bounds $\overline{p}_{\Gamma}$ become larger and will eventually
converge to $1$ as $\Gamma \to \infty$.

\begin{table}[t]
  \centering
  \caption{Illustration of a two-sided sensitivity analysis table and
    the corresponding sensitivity values.}
  \label{tab:gene}
  \begin{tabular}{|l|lrrrrrr|rr|}
    \hline
    probe set & \multicolumn{7}{c|}{sensitivity analysis} & \multicolumn{2}{c|}{sensitivity
      value} \\
    \hline
    \multirow{2}{*}{41214\_at} & $\Gamma$ & 1 & 2 & 3 & 5 &
    7 & 10 & {\bf 4.69} & {\bf 8.10} \\
    & $\overline{p}_{\Gamma}$ & 0.00 & 0.00 & 0.00 & 0.01 & 0.03 & 0.08
    & 0.01 & 0.05 \\
    \hline
    \multirow{2}{*}{38355\_at} & $\Gamma$ & 1 & 2 & 3 & 5 &
    7 & 10 & {\bf 4.69} & {\bf 8.10} \\
    & $\overline{p}_{\Gamma}$ & 0.00 & 0.00 & 0.00 & 0.01 & 0.03 & 0.08
    & 0.01 & 0.05 \\
    \hline
    \multirow{2}{*}{37583\_at} & $\Gamma$ &1 & 2 & 3 & 5 &
    7 & 10 & {\bf 1.84} & {\bf 2.44} \\
    & $\overline{p}_{\Gamma}$ & 0.00 & 0.02 & 0.13 & 0.60 & 1.00 & 1.00
    & 0.01 & 0.05 \\
    \hline
    \multirow{2}{*}{35885\_at} & $\Gamma$ &1 & 2 & 3 & 5 &
    7 & 10 & {\bf 1.79} & {\bf 2.36} \\
    & $\overline{p}_{\Gamma}$ & 0.00 & 0.02 & 0.15 & 0.66 & 1.00 & 1.00
    & 0.01 & 0.05 \\
    \hline
    \multirow{2}{*}{32052\_at} & $\Gamma$ &1 & 2 & 3 & 5 &
    7 & 10 & {\bf 1.68} & {\bf 2.20} \\
    & $\overline{p}_{\Gamma}$ & 0.00 & 0.03 & 0.20 & 0.80 & 1.00 & 1.00
    & 0.01 & 0.05 \\
    \hline
    \multirow{2}{*}{34477\_at} & $\Gamma$ &1 & 2 & 3 & 5 &
    7 & 10 & {\bf 1.50} & {\bf 1.93} \\
    & $\overline{p}_{\Gamma}$ & 0.00 & 0.06 & 0.33 & 1.00 & 1.00 & 1.00
    & 0.01 & 0.05 \\
    \hline
    \multirow{2}{*}{38446\_at} & $\Gamma$ &1 & 2 & 3 & 5 &
    7 & 10 & {\bf 1.43} & {\bf 1.84} \\
    & $\overline{p}_{\Gamma}$ & 0.00 & 0.08 & 0.40 & 1.00 & 1.00 & 1.00
    & 0.01 & 0.05 \\
    \hline
    \multirow{2}{*}{31687\_f\_at} & $\Gamma$ &1 & 2 & 3 & 5
    & 7 & 10 & {\bf 1.24} & {\bf 1.57} \\
    & $\overline{p}_{\Gamma}$ & 0.00 & 0.17 & 0.67 & 1.00 & 1.00 & 1.00
    & 0.01 & 0.05 \\
    \hline
    \multirow{2}{*}{31525\_s\_at} & $\Gamma$ &1 & 2 & 3 & 5
    & 7 & 10 & {\bf 1.23} & {\bf 1.57} \\
    & $\overline{p}_{\Gamma}$ & 0.00 & 0.17 & 0.68 & 1.00 & 1.00 & 1.00
    & 0.01 & 0.05 \\
    \hline
  \end{tabular}
\end{table}

When there are many hypotheses tested at the same time, the full
sensitivity analysis produces a lengthy table (the middle columns of
\Cref{tab:gene} under ``sensitivity analysis'') and is a rather
inefficient way of presenting information. In this paper we propose a new
quantity---\emph{sensitivity value}---to summarize the sensitivity analyses. The sensitivity
value is simply the critical $\Gamma$ where the $p$-value upper bound
$\overline{p}_{\Gamma}$ crosses a pre-specified significance level
$\alpha$; for the formal definition, see \Cref{sec:defin-sens-value}. This
concept is illustrated in the last two columns of
\Cref{tab:gene}, where the bolded numbers are the corresponding
sensitivity values of the probe sets.

Although the term ``sensitivity value'' is new, it has already been
routinely reported in observational studies to strengthen their
qualitative conclusions.
The sensitivity value speaks to the assertion ``it might be bias'' in an
observational study in much the same way as the $p$-value speaks to the
assertion ``it might be bad luck'' in a randomized trial \citep[Section
1.2]{rosenbaum2015two}. A large sensitivity value means that it would
take a large bias (departure from random assignment) for an
association between treatment and outcome to be non-causal in an
observational study, just as a small $p$-value in a randomized trial
means it would take a large amount of bad luck for the association
to be due to chance alone. See
\Cref{sec:discussion} for more discussion on the different roles of
$p$-value and sensitivity value.

The main goal of this paper is to investigate how to design an
observational study to maximize its sensitivity value (in a stochastic
sense). Previously,
this objective is indirectly pursued by maximizing the probability
that the $p$-value upper bound is less than $\alpha$ at a fixed sensitivity
level $\Gamma$
\citep{heller2009split,rosenbaum2010design,rosenbaum2015bahadur}. This paper takes the
first step towards directly achieving this goal by establishing the
asymptotic distribution of the sensitivity value in pair-matched
studies. Additionally, we explore the potential to use sensitivity
values in genomics screening when unobserved confounding is a major
concern.

In \Cref{sec:revi-sens-analys} we review sensitivity analysis for
pair-matched observational study. We formally define sensitivity value
in \Cref{sec:defin-sens-value} and derive
its asymptotic distribution in \Cref{sec:main-result}. Then in
\Cref{sec:selecting-statistics,sec:select-subp,sec:selecting-outcomes}
we discuss the implications of our theoretical results in designing
observational studies. We conclude the paper with some brief
discussion in \Cref{sec:discussion}. Technical proofs can be found in
the supplementary file.

\section{Review: Sensitivity Analysis}
\label{sec:revi-sens-analys}

Consider a typical setting of an observational study with $I$
independent matched pairs,
$i=1,\dotsc,I$. Each pair has two subjects, $j=1,2$, one
treated, denoted by $Z_{ij} = 1$, and one control, denoted by $Z_{ij} =
0$. Pairs are matched for observed covariates so $x_{i1} = x_{i2}$, but the investigator
may be concerned that matching failed to control for an unmeasured
confounder $u_{ij}$, so possibly $u_{i1} \ne u_{i2}$ for some or all
$i$.  Let $r_{Tij}$ be the potential outcome of the $j$-th
subject in the $i$-the pair if subject $j$ in matched pair $i$ receives treatment. Similarly,
$r_{Cij}$ is the potential outcome if the subject receives control. The
observed outcome is $R_{ij} = Z_{ij} r_{Tij} + (1 - Z_{ij}) r_{Cij}$
and the individual treatment effect $r_{Tij} - r_{Cij}$ cannot be
observed for any subject
\citep{rubin1974estimating}. Let  $Y_i$ be the treatment-minus-control difference $Y_i = (Z_{i1} -
Z_{i2})(R_{i1} - R_{i2})$ for the $i$-the pair. Let $\mathcal{F} =
\{(r_{Tij},r_{Cij},x_{ij},u_{ij}),~i=1,\dots,I,~j=1,2 \}$ and
$\mathcal{Z}$ be the event that $\{Z_{i1} + Z_{i2} =
1,~i=1,\dotsc,I\}$. The
sharp null hypothesis of no treatment effect assumes that $H_0:r_{Tij}
= r_{Cij},~\forall i,j$. If $H_0$ is true and the treatments are
randomly assigned (i.e.\ $\mathrm{P}(Z_{i1}=1|\mathcal{F},\mathcal{Z})
= 1/2$ for all $i$), then conditioning on $\mathcal{F}$ and
$\mathcal{Z}$, $Y_i = (Z_{i1} -
Z_{i2})(r_{Ci1} - r_{Ci2})$ attaches equal probabilities to $\pm
|r_{Ci1} - r_{Ci2}|$.

To test for $H_0$, a commonly used family of statistics are the signed
score statistics
\begin{equation}
  \label{eq:signed-score}
  T(Z,R) = \frac{\sum_{i=1}^I \mathrm{sgn}(Y_i) q_i}{\sum_{i=1}^I q_i},
\end{equation}
where $\mathrm{sgn}(y) = 1_{y > 0}$ and $q_i \ge 0$ is a function of $|Y_i|$
such that $q_i = 0$ if $Y_i = 0$. A special case is Wilcoxon's signed
rank statistic for which $q_i = \mathrm{rank}(|Y_i|)$. The statistic $T$
in \eqref{eq:signed-score} is normalized by $\sum_{i=1}^I q_i$ so it
is always between $0$ and $1$. Under $H_0$ and random treatment
assignment, conditioning on $\mathcal{F}$ and $\mathcal{Z}$, $q_i$ are
fixed constants and $\mathrm{sgn}(Y_i)$ are i.i.d.\ Bernoulli
variables with
$\mathrm{P}(\mathrm{sgn}(Y_i) = 1) = \mathrm{P}(\mathrm{sgn}(Y_i) = 0)
= 1/2$. This yields the null distribution of signed score
statistics. The exact distribution is usually difficult to compute for
large $I$. In this case, Monte-Carlo simulations or central limit
theorems can be used to approximate the distribution of $T$.

In an observational study, matching may fail to control a relevant
unobserved covariate $u_{ij}$, so $\mathrm{P}(Z_{i1} = 1|\mathcal{F}) \ne
1/2$. A simple model for sensitivity analysis in an observational
study asserts that the odds of treatment deviates from $1$ by at most
a factor of $\Gamma \ge 1$,
\begin{equation}
  \label{eq:sensitivity-model}
  \frac{1}{\Gamma} \le \frac{\mathrm{P}(Z_{i1} =
    1|\mathcal{F},\mathcal{Z})}{\mathrm{P}(Z_{i1} = 0|\mathcal{F},\mathcal{Z})} \le \Gamma,~i=1,\dotsc,I,
\end{equation}
with independent assignments in distinct pairs. $\Gamma = 1$ yields
random assignment and each fixed $\Gamma > 1$ indicates an unknown but
limited departure from random assignment.

A typical sensitivity analysis computes the range of plausible
$p$-values using the test statistic \eqref{eq:signed-score} under
the sensitivity model \eqref{eq:sensitivity-model}. Let
$\overline{T}_{\Gamma}$ be the sum of $I$ independent random
variables, $i=1,\dotsc,I$, taking the value $q_i$ with probability
$\Gamma/(1+\Gamma)$ and $0$ with probability
$1/(1+\Gamma)$. Note that this is also well-defined for $0 < \Gamma <
1$. Similarly, let $\underline{T}_{\Gamma}$ be the random variable created
by replacing $\Gamma$ with $1/\Gamma$ in the definition of
$\overline{T}_{\Gamma}$. In other words, $\underline{T}_{\Gamma}
\overset{d}{=} \overline{T}_{1/\Gamma}$. This fact will be useful when
we define the untruncated sensitivity value in the next Section.

In \citet[Section 4.4]{rosenbaum2002observational},
it is shown that, under the sensitivity model \eqref{eq:sensitivity-model},
\begin{equation} \label{eq:rosenbaum-bound}
  \underline{p}_{\Gamma} = \mathrm{P}(\underline{T}_{\Gamma} \ge t|\mathcal{F},\mathcal{Z}) \le
  \mathrm{P}(T \ge t|\mathcal{F},\mathcal{Z}) \le
  \mathrm{P}(\overline{T}_{\Gamma} \ge
  t|\mathcal{F},\mathcal{Z}) = \overline{p}_{\Gamma},~\forall t,~\Gamma \ge 1.
\end{equation}
The bounds are sharp in the sense that they can be attained for
a particular $\mathrm{P}(Z_{i1}=1|\mathcal{F},\mathcal{Z})$ satisfying
\eqref{eq:sensitivity-model}. When $\Gamma = 1$ (no unmeasured
confounder), both bounding distributions are equal to
the null distribution of $T$. Therefore $\overline{p}_1 =
\underline{p}_1$ are equal to the conventional $p$-value.

When the sample size $I$ is large, the distribution of the
bounding variable $\overline{T}_{\Gamma}$ can be approximated by a central limit
theorem \citep[Section 6.1]{sidak1999theory}. Conditioning on
$\mathcal{F}$ and $\mathcal{Z}$, we have
\begin{equation} \label{eq:t-gamma}
  \sqrt{I} \cdot \frac{\overline{T}_{\Gamma} - \Gamma/(1 + \Gamma)}{\sqrt{\Gamma/(1 + \Gamma)^2 \sigma_{q,I}^2}} \overset{d}{\to}
  \mathrm{N}(0, 1),~\mathrm{where}~
  \sigma_{q,I}^2 = \frac{I^{-1}\sum_{i=1}^I q_i^2}{\big(I^{-1}\sum_{i=1}^I q_i\big)^2},
\end{equation}
providing $(\sum_{i=1}^I q_i^2) / (\max_i q_i^2) \to \infty$. In this
paper we further assume $\lim_{I \to \infty} \sigma_{q,I}^2 =
\sigma_q^2$ exists. This holds for Wilcoxon's signed rank test and all
other test statistics considered in this paper. The $p$-value upper
bound $\overline{p}_{\Gamma}$ can be subsequently approximated by the
tail probability of the normal distribution.

\section{Definition of sensitivity value}
\label{sec:defin-sens-value}

We are ready to give the formal definition of sensitivity value:
\begin{definition} \label{def:sen-val}
  Given the data $(Z,R)$ and significance level $\alpha$, the
  \emph{truncated sensitivity value} is the smallest $\Gamma \ge 1$ such that
  the upper bound $\overline{p}_{\Gamma}(Z,R)$ is not significant. Formally,
  \begin{equation} \label{eq:gamma-star-star-definition}
    \Gamma_{\alpha}^{**}(Z,R) = \inf\Big\{\Gamma \ge 1\,|\,
    \overline{p}_{\Gamma}(Z,R) > \alpha
    \Big\}.
  \end{equation}
\end{definition}

Note that the $p$-value upper bound $\overline{p}_{\Gamma}$ is always
increasing in $\Gamma$, so
the set in \eqref{eq:gamma-star-star-definition} is an interval
$[\Gamma_{\alpha}^{**},\infty)$ and its infimum is well defined.
Note that, similar to Fisher's $p$-value, the sensitivity value is a
deterministic function of the data.

By definition, $\Gamma_{\alpha}^{**} = 1$ if $\overline{p}_1 \ge \alpha$ (the
naive $p$-value is not significant). Therefore, when
analyzing the distribution of $\Gamma_{\alpha}^{**}$ with randomly generated
data $(Z,R)$, $\Gamma_{\alpha}^{**}$ usually has a point mass at $1$.
We find it more convenient to consider the untruncated version of
$\Gamma_{\alpha}^{**}$,
\begin{equation} \label{eq:gamma-star-definition}
  \Gamma_{\alpha}^{*}(Z,R) = \inf\Big\{\Gamma > 0\,|\,
  \overline{p}_{\Gamma}(Z,R) \ge \alpha
  \Big\}.
\end{equation}
Although a typical sensitivity analysis is only performed for $\Gamma \ge 1$, the
bounding variable $\overline{p}_{\Gamma}$ can still be defined for $0 <
\Gamma < 1$ and it is obvious that $\Gamma_{\alpha}^{**} =
\max(\Gamma_{\alpha}^{*}, 1)$. By allowing the sensitivity value to
be less than $1$, $\Gamma_{\alpha}^{*}$ becomes a continuous variable
and provides extra information. To see
this, if the $p$-value is not significant under $\Gamma = 1$, the
untruncated sensitivity value $\Gamma_{\alpha}^{*} < 1$ and
its reciprocal $1 / \Gamma_{\alpha}^{*}$ is
where the $p$-value lower bound $\underline{p}_{\Gamma}$ first becomes
significant, since $\underline{p}_{1/\Gamma} = \overline{p}_{\Gamma}$
(see the paragraph before
equation \eqref{eq:rosenbaum-bound}). In other words, when $\Gamma_{\alpha}^{*}
< 1$, $1/\Gamma_{\alpha}^{*}$ is
the smallest magnitude of bias needed to make the test significant. In
the development below we will always work with $\Gamma_{\alpha}^{*}$ and
refer to it as \emph{sensitivity value}.

To compute the sensitivity value, we can use the normal approximation
of the bounding variable $\bar{T}_{\Gamma}$ in
\eqref{eq:t-gamma}. In what follows, we will suppress $\alpha$ in the subscript
of $\Gamma^{*}_{\alpha}$ if it causes no confusion. Let $\kappa^{*} =
\Gamma^{*} / (1 + \Gamma^{*})$ so $\Gamma^{*} =
1$ corresponds to $\kappa^{*} = 1/2$. The value $\kappa^{*}$, referred
to as \emph{transformed sensitivity value} hereafter, should solve
$
\sqrt{I} \cdot (T - \kappa) = \sqrt{\kappa (1 - \kappa)} \sigma_q
\cdot \big(\bar{\Phi}^{-1}( \alpha) + o_p(1) \big)
$
where $\bar{\Phi}^{-1}(\alpha)$ is the
upper-$\alpha$ quantile of the standard normal distribution. Taking
the square of this equation and then solving a quadratic equation
of $\kappa^{*}$, we obtain
\begin{equation} \label{eq:kappa-star-t}
  \kappa^{*} = \frac{2IT + c^2 -
    \sqrt{4c^2 I T(1-T) + c^4}}{2(I + c^2)} +
  o_p\Big(\frac{1}{\sqrt{I}}\Big)
  ,
\end{equation}
where $c =\sigma_{q,I} \bar{\Phi}^{-1}( \alpha)$.
The larger root is discarded because $T - \kappa^{*}$ must be
non-negative. The sensitivity value $\Gamma^{*}$ can be subsequently
obtained by $\Gamma^{*} = \kappa^{*} / (1 - \kappa^{*})$.
As a remark, the additional $o_p(1/\sqrt{I})$ term in \eqref{eq:kappa-star-t}
comes purely from the normal approximation of the bounding variable
$\bar{T}_{\Gamma}$ in \eqref{eq:t-gamma}. The normal approximation is
known to be very accurate for Wilcoxon's signed rank test for as small
as $30$ matched pairs.

Alternatively, the exact sensitivity value
may be computed by binary-searching a full sensitivity analysis as
demonstrated in \Cref{tab:gene}. This method is free of asymptotic
error but more computationally intensive.  \Cref{tab:accuracy} reports
the difference between the transformed sensitivity values $\kappa^{*}$
computed by the approximation \eqref{eq:kappa-star-t} and by
grid-searching a full sensitivity analysis table. Since the exact
distribution of $\overline{T}_{\Gamma}$ is too complicated even for
moderate sample size, we approximate it using $10^{5}$ Monte-Carlo
samples. In most cases, the asymptotic approximation
\eqref{eq:kappa-star-t} is quite accurate, especially if the sample
size $I$ or the significance level $\alpha$ is not too small.

\begin{table}[t]
  \centering
  \caption{Accuracy of formula \eqref{eq:kappa-star-t} for computing
    sensitivity value. In each scenario, we compute two approximations
    of the sensitivity value: the transformed sensitivity value
    $\kappa^{*}$ computed by formula \eqref{eq:kappa-star-t}, and a
    finite-sample value computed by grid-searching
    a Monte-Carlo sensitivity analysis table (at each $\Gamma$ we
    compute the $p$-value upper bound $\overline{p}_{\Gamma}$ by
    $100,000$ realizations of $\overline{T}_{\Gamma}$). This table reports the
    mean and $10\%$ and $90\%$ quantiles in
    $100$ simulations of the differences between the two
    approximations of $\kappa^{*}$.}
  \label{tab:accuracy}
  \begin{tabular}{|ll|ccc|ccc|}
    \hline
    & & \multicolumn{3}{c|}{$\alpha = 0.05$} & \multicolumn{3}{c|}{$\alpha = 0.005$} \\
    I & dist.\ of $Y$ & 10\% & mean & 90\% & 10\% & mean & 90\% \\
    \hline
    30 & $\mathrm{N}(1, 1)$  & $0.006$ & $0.010$ & $0.015$ & $0.018$ & $0.031$ & $0.047$ \\
    & $t_2+1.5$  & $0.003$ & $0.009$ & $0.015$ & $0.011$ & $0.030$ & $0.050$ \\
    100 & $\mathrm{N}(1, 1)$  & $0.004$ & $0.004$ & $0.005$ & $0.012$ & $0.014$ & $0.016$ \\
    & $t_2+1.5$  & $0.003$ & $0.004$ & $0.005$ & $0.008$ & $0.012$ & $0.014$ \\
    \hline
  \end{tabular}
\end{table}

So far we have only discussed sensitivity analysis for one-sided
test. Following the suggestion by \citet[Section 4.2]{cox1977role}, a
simple way to obtain a two-sided $p$-value in sensitivity analysis
(e.g.\ in \Cref{tab:gene}) is
to double the smaller of the two one-sided $p$-value upper bounds.
Consequently, to compute the
two-sided sensitivity value, one can simply take the maximum of the two
one-sided sensitivity values with significance level $\alpha / 2$.

\section{Distribution of sensitivity value}
\label{sec:main-result}

\subsection{Asymptotic normality}
\label{sec:asymptotic-normality}

Next we derive the asymptotic
distribution of the sensitivity value $\Gamma^{*}$ when the data
$(Z_i,R_i)$ are generated i.i.d.\ from $F$. We should emphasize that
all our theoretical analysis is made in the favorable situation that
$F$ satisfies the random treatment assignment
mechanism $\mathrm{P}(Z_{i1}=1|\mathcal{F},\mathcal{Z}) = 1/2$ but possibly
has a non-zero treatment effect \citep{rosenbaum2010design}.

Under weak regularity
conditions, the test statistic $T$ has a normal limiting distribution
\citep[Section 2.8]{hettmansperger1984statistical}:
\begin{equation} \label{eq:alter-dist}
  \sqrt{I} \cdot \frac{T - \mu_F}{\sigma_F} \overset{d}{\to}
  \mathrm{N}(0, 1).
\end{equation}
The mean and variance parameters usually depend on the distribution
$F$. When $F$ satisfies the null hypothesis of no treatment effect,
then $\mu_F = 1/2$ and $\sigma_F^2 = \sigma_q^2/4$.

Combining the approximation \eqref{eq:alter-dist} and formula
\eqref{eq:kappa-star-t}, after some algebra we get
\begin{theorem} \label{thm:kappa-star}
  Assume the central limit theorem \eqref{eq:t-gamma} holds for the bounding variable
  $\overline{T}_{\Gamma}$ and $\lim_{I \to \infty} \sigma_{q,I}^2 =
  \sigma_q^2$ exists. For test statistic $T$ satisfying
  \eqref{eq:alter-dist}, the transformed
  sensitivity value $\kappa^{*}_{\alpha}$ for fixed $0 < \alpha < 1$ has an asymptotic normal distribution:
  \begin{equation}
    \label{eq:kappa-star}
    \sqrt{I} \cdot \Big[\kappa_{\alpha}^{*} - \mu_F
    \Big] \overset{d}{\to} \mathrm{N}\bigg(-\sigma_q \bar{\Phi}^{-1}(\alpha)
    \sqrt{\mu_F(1 - \mu_F)}, ~\sigma_F^2\bigg),
  \end{equation}
  where $\mathrm{N}$ is the standard normal distribution and
  $\bar{\Phi}(y) = 1 - \Phi(y)$ is its complementary CDF.
\end{theorem}


The value $\mu_F$, or more precisely the corresponding sensitivity
value $\tilde{\Gamma} = \mu_F / (1 - \mu_F)$, is called ``design sensitivity'' by
\citet{rosenbaum2004design}. This value describes how sensitive a test
statistic $T$ is to unobserved bias when the sample size $I \to \infty$. In other
words, the design sensitivity $\tilde{\Gamma}$ is the (stochastic)
upper bound of the sensitivity value $\Gamma^{*}$.
When the distribution $F$ satisfies Fisher's sharp null,
the design sensitivity $\tilde{\Gamma}$ is $1$.

In \Cref{thm:kappa-star} we assume the significance level $\alpha$ is
fixed. This is useful to eliminate several terms in
\eqref{eq:kappa-star-t}. In finite samples, the ratio $c^2/I$ can be
nonnegligible when $I$ is moderate. For example, when $\alpha = 0.05$
and Wilcoxon's signed rank test is used, $c^2 \approx 3.6$ and $c^2/I$
is nonnegligible for $I = 50$. If we assume $\sigma_q^2
\bar{\Phi}^{-1}(\alpha)^2 / I \approx \eta > 0$, then using an
asymptotic analysis similar to \Cref{thm:kappa-star}, we have
\begin{equation}
  \label{eq:kappa-star-gamma}
  \begin{split}
    \sqrt{I} \bigg[\kappa_{\alpha}^{*} - \Big(\mu_F - \frac{(2 \mu_F - 1) \eta +
      \sqrt{4\eta \mu_F(1-\mu_F) +
        \eta^2}}{2(1+\eta)}\Big)\bigg] \\
    \approx \mathrm{N}\bigg(0,
    \frac{\sigma_F^2}{(1+\eta)^2} \Big( 1 +
    \frac{\eta(2\mu_F-1)}{\sqrt{4\eta \mu_F(1-\mu_F) + \eta^2}} \Big)^2
    \bigg).
  \end{split}
\end{equation}
The relationship in \eqref{eq:kappa-star-gamma} is not convergence in
distribution, because to make $c^2/I$ converging to a constant, the
significance level $\alpha$ has to decrease to $0$ and the normal
approximation \eqref{eq:t-gamma} of $\overline{T}_{\Gamma}$ becomes
less and less accurate.
Nonetheless, we find \eqref{eq:kappa-star-gamma} provides a
more accurate approximation of $\kappa^{*}_{\alpha}$ than
\eqref{eq:kappa-star} when sample size is moderate and $\alpha$ is not
too small.

In a related work, \citet{rosenbaum2015bahadur} derived the limit of
$\log \overline{p}_{\Gamma}$ using large deviations
theory. The asymptotic results in
\Cref{thm:kappa-star}, the finite sample approximation
\eqref{eq:kappa-star-gamma}, and the approximation in
\citet{rosenbaum2015bahadur} should be used for different
purposes. \Cref{thm:kappa-star} describes the asymptotic behavior of
the transformed sensitivity value and in particular how
$\Gamma^{*}$ converges to the design sensitivity
$\tilde{\Gamma}$. Equation \eqref{eq:kappa-star-gamma} is more accurate in
computing the power of sensitivity analysis when sample size $I$ is
moderate. The large deviations approximation in
\citet{rosenbaum2015bahadur} can be inverted to approximate the
sensitivity value, but it is applicable only if the significance level
$\alpha$ is very small and is more difficult to compute as it uses
the moment generating function of $F$ rather just the first two
moments.


\subsection{Power of sensitivity analysis}
\label{sec:power-sens-analys}

\citet{rosenbaum2004design} defines the \emph{power} of a sensitivity
analysis as the probability that the test rejects the null hypothesis
at sensitivity level $\Gamma$ \citep[see also][Chapter
14]{heller2009split,rosenbaum2010design}. Let $c_{\Gamma,\alpha}$ be
the upper $\alpha$-quantile of ${\overline{T}}_{\Gamma}(Z, R)$. Then
using \Cref{thm:kappa-star},
the power at sensitivity level $\Gamma=\kappa/(1-\kappa)$ for some
$\kappa < \mu_F$ is given by
\begin{equation} \label{eq:power}
  \begin{split}
    \mathrm{P}(T(Z,R) \ge c_{\Gamma,\alpha}) &=
    \mathrm{P}(\kappa_{\alpha}^{*} > \kappa)
    \approx \Phi\left(\frac{\sqrt{I}(\mu_F - \kappa) - \sigma_q \bar{\Phi}^{-1}( \alpha)
        \sqrt{\mu_F(1 - \mu_F)}}{\sigma_F}\right).
  \end{split}
\end{equation}

\citet[equation (4)]{heller2009split} derived a similar formula
without specifying the constant term $\sigma_q
\bar{\Phi}^{-1}( \alpha) \sqrt{\mu_F(1-\mu_F)}$ in
\eqref{eq:power}. Because \citet{heller2009split} intended to
derive the design sensitivity $\tilde{\Gamma}$ (the asymptotic limit
of $\kappa_{\alpha}^{*}$), it is unnecessary for them to compute the
constant term exactly. However, the constant term can be substantial
in power approximation if the target sensitivity level $\kappa$ is
close to $\mu_F$. Alternatively, we can get a more accurate power
approximation using the finite-sample approximation
\eqref{eq:kappa-star-gamma}.

As an example, consider the Wilcoxon signed rank statistic which
corresponds to $q_i =
\mathrm{rank}(|Y_i|)$. In this case, $\sum_{i} q_i = I(I+1)/2$ and
$\sum_{i} q_i^2 = I(I+1)(2I+1) / 6$, hence $\sigma_{q,I}^2 \to
4/3$. \citet[Section 2.5]{hettmansperger1984statistical} showed that
$\mu_F = \mathrm{P}(Y_1 + Y_2 > 0)$ and $\sigma_F^2 = 4[\mathrm{P}(Y_1 +
Y_2 > 0) - \mathrm{P}(Y_1 + Y_2 > 0, Y_1 + Y_3 > 0)^2]$. Suppose
$Y \sim \mathrm{N}(0.5, 1)$ and the sample size is $I =
200$. Wilcoxon's test has mean
$\mu_F = \Phi(1/\sqrt{2}) \approx 0.76$ (corresponds to design
sensitivity $\tilde{\Gamma} = \mu_F/(1-\mu_F) \approx 3.17$) and the
variance $\sigma_F^2$ is about $0.26$. Suppose we are interested the power at
$\Gamma = 2.5$ and $\alpha = 0.05$. Using $10,000$ simulations, we find that the actual
power is about $33.6\%$. The approximate power using
\eqref{eq:kappa-star} is $37.1\%$, the approximate power using
\eqref{eq:kappa-star-gamma} is $33.5\%$, and the power calculated
ignoring the constant term in \eqref{eq:kappa-star} is $90.8\%$.

\section{Selecting test statistics}
\label{sec:selecting-statistics}

In the next three Sections, we discuss the implications of the results
obtained in \Cref{sec:main-result} in selecting the design of an
observational study. First, we consider how to maximize the
sensitivity value $\Gamma^{*}$ by picking a test statistic.

Consider the general signed score statistic \eqref{eq:signed-score} with $q_i=
\psi\big(\mathrm{rank}(|Y_i|)/(I+1)\big)$, where the function $\psi(u)
\ge 0,~0 < u < 1$ satisfies $\int_0^1 \psi(u) \diff u < \infty$, and $\int_0^1
\psi^2(u) \diff u < \infty$. Then
\[
\sigma^2_{q,I} = \frac{(1/I)\sum_{i=1}^I q_i^2}{\big[(1/I)\sum_{i=1}^I
  q_i\big]^2} \to \frac{\int_0^1 \psi^2(u) \diff u}{\Big(\int_0^1
  \psi(u) \diff u\Big)^2} = \frac{\|\psi\|_2^2}{\|\psi\|_1^2}.
\]

Under the alternative model that $Y_i \overset{\mathrm{i.i.d.}}{\sim} F$,
the asymptotic distribution of $T(Z,R)$ is given by the normal approximation
\eqref{eq:alter-dist} with mean \citep[page 104]{hettmansperger1984statistical}
\begin{equation} \label{eq:mu-F-psi}
\mu_F = \mu_F[\psi] = \frac{\int_0^\infty \psi(\mathrm{P}(|Y| \le y)) \diff
  F(y)}{\int_0^1 \psi(u)} = \frac{\langle \psi, g \rangle}{\|\psi\|_1},
\end{equation}
where $\langle \psi, g \rangle = \int_0^1 \psi(u)g(u)\diff u$ and
\[
g(u) = \frac{f((F^{+})^{-1}(u))}{f((F^{+})^{-1}(u)) + f(-(F^{+})^{-1}(u))},~F^{+}(y) =
\mathrm{P}(|Y| \le y) = F(y) - F(-y),~y>0.
\]
The variance parameter $\sigma_F^2$ is more complicated
and in our theoretical analysis we will only compare the means of
$\kappa^{*}$ for different statistics.

\begin{figure}[t]
  \centering
  \begin{subfigure}[t]{0.48\textwidth}
    \includegraphics[width = \textwidth]{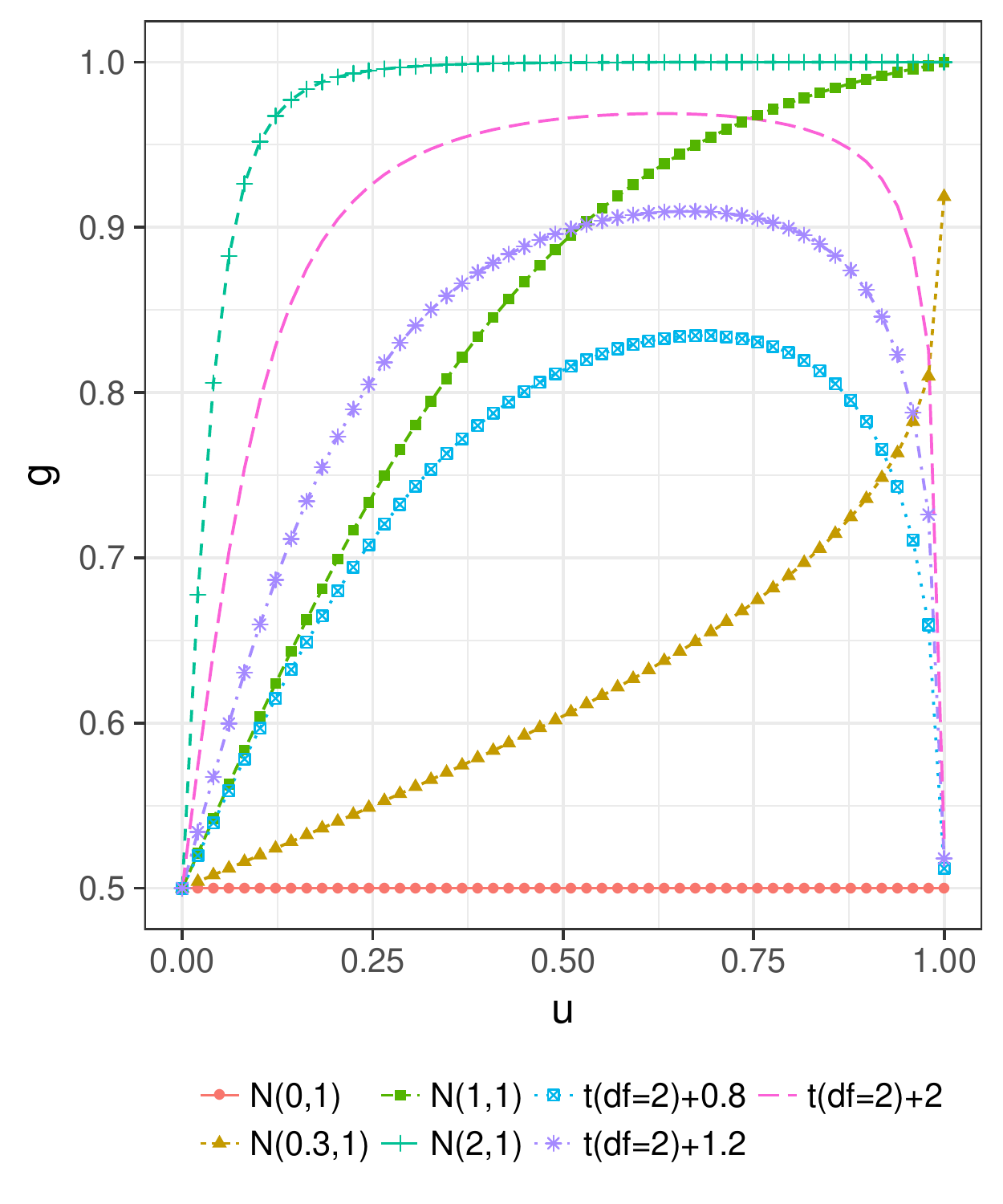}
    \caption{Function $g(u)$ for design sensitivity. The location shift
      is $0$, $0.5$, $1$ or $2$ and the noise distribution is normal or
      $t$-distribution with $2$ degrees of freedom.}
    \label{fig:g}
  \end{subfigure} \quad
  \begin{subfigure}[t]{0.48\textwidth}
    \includegraphics[width = \textwidth]{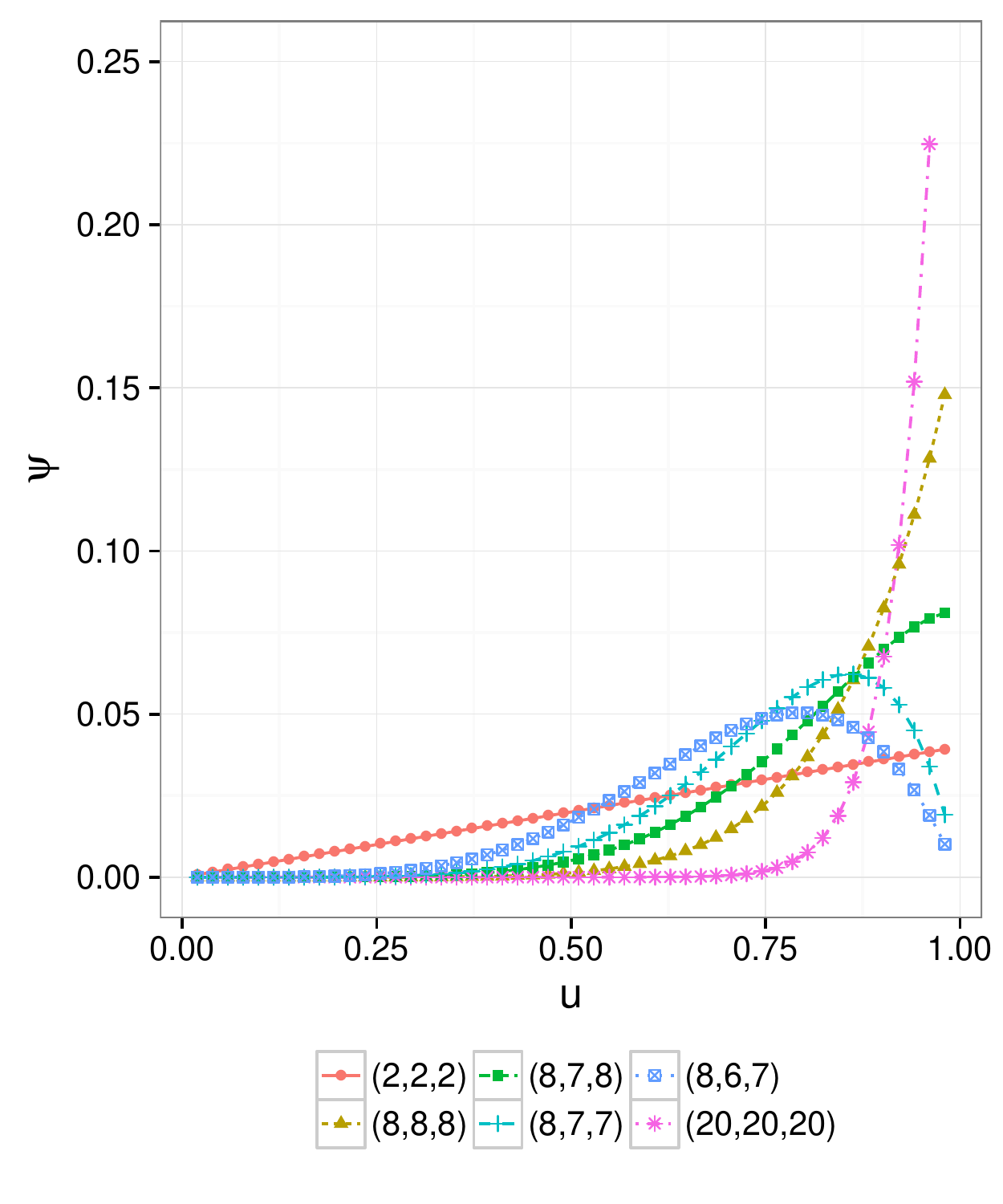}
    \caption{Score functions $\psi(u)/\|\psi\|_1$ for several
      U-statistics.}
    \label{fig:psi}
  \end{subfigure}

  \caption{Design sensitivity is determined by the inner product of
    $g$ (left plot) and $\psi/\|\psi\|_1$ (right plot).}

\end{figure}

By \Cref{thm:kappa-star}, $\mu_F$ is the limit of the transformed sensitivity
value $\kappa^{*}$ when the sample size $I \to \infty$. Formula \eqref{eq:mu-F-psi}
suggests that $\mu_F$ is just a weighted average of $g(u)$. Notice
that $g(0) = 1/2$, $g(u) \le
1$, and $\lim_{u \to 1} g(u)$ depends on the tail of $F$. To see this,
suppose $F$ is symmetric and unimodal with mean $d$. Furthermore
assume the density $f(y) = f_0(y-d)$ is positive on
the real line, so $(F^{+})^{-1}(u) \to \infty$ as $u \to 1$. Then
\[
\lim_{u \to 1} g(u) = \lim_{y \to \infty} \frac{1}{1 + f_0(y-d)/f_0(-y-d)}.
\]
Therefore, if the tail is a power law, $f_0(y) \propto
|y|^{-\lambda}$, then $\lim_{u \to 1} g(u) = 1/2$ and hence $g(u)$
cannot be monotonically increasing. If the tail decay is exponential,
$f_0(y) \propto e^{-\lambda |y|}$, then $\lim_{u \to 1} g(u) \in
(1/2,1)$. If the tail decay is faster than exponential, for example
$f_0(y) \propto e^{-\lambda y^2}$, then $\lim_{u \to 1} g(u) = 1$.
\Cref{fig:g} plots the function $g(u)$ for some familiar
distributions. 

\citet{rosenbaum2004design} proposed to select the test statistics to
maximize the design sensitivity $\tilde{\Gamma}$ or equivalently
$\mu_F$ in the transformed scale; see also
\citet{rosenbaum2011new,rosenbaum2012design}. With this objective in
mind, the optimal choice of
the score function $\psi$ should converge to $\delta_{u^{*}}$, where
$\delta_u$ is the Dirac-$\delta$ function and $u^{*} = \argmax_u
g(u)$.

However, in finite samples $I < \infty$, \Cref{thm:kappa-star}
suggests that the mean of $\kappa^{*}$ is approximately
\[
\mu_{F,I}[\psi] = \langle \psi, g \rangle -
\frac{\bar{\Phi}^{-1}(\alpha)}{\sqrt{I}} \|\psi\|_2 \sqrt{\langle \psi, g \rangle (1 - \langle \psi, g \rangle)}
\]
provided that $\psi$ is normalized so that $\|\psi\|_1 =
1$. Therefore, it is not a good idea to choose a spiky score function
$\psi$ since the $L_2$ norm of $\psi$ can blow up to infinity.


As an example, consider the following class of U-statistics proposed by
\citet{rosenbaum2011new} that are indexed by three parameters
$(m,\underline{m},\overline{m})$. Let $h(y)$ be a function of $m$
variables that count the number of positive differences among
the order statistics between $|y|_{(\underline{m})}$ and
$|y|_{(\overline{m})}$. The corresponding U-statistic is defined as
$
T = {I \choose m}^{-1} \sum_{|\mathcal{I}|=m} h(Y_{\mathcal{I}}),
$
which can be written as the signed score form
\eqref{eq:signed-score}. Let $a_i$ be the rank of $|Y_i|$. In absence
of ties, the score of $Y_i$ is given by \citep[Section 3.1]{rosenbaum2011new}
\begin{equation*} \label{eq:u-score}
  q_i = {I \choose m}^{-1} \sum_{l = \underline{m}}^{\overline{m}} {{a_i
      - 1} \choose {l-1}} {{I - a_i} \choose {m-l}} \approx I^{-1}
  \sum_{l=\underline{m}}^{\overline{m}} l {m \choose l} p^{l-1}
  (1-p)^{m-l}~\mathrm{for}~p=\mathrm{rank}(|Y_i|)/I.
\end{equation*}
Note that the choice $(m,\underline{m},\overline{m})=(2,2,2)$
closely approximates Wilcoxon's statistic
\citep{rosenbaum2011new}. \Cref{fig:psi}
plots the approximate score function $\psi$ for several choices of
$(m,\underline{m},\overline{m})$.


Tables \ref{tab:u1} and \ref{tab:u2} show the median and standard deviation of
$\kappa^{*}$ using various U-statistics. Our theoretical
approximations are very close to the values obtained by simulations.
When $Y \sim \mathrm{N}(0.3, 1)$, the design sensitivity is maximized
by $(20,20,20)$. This statistic still has the
largest mean when $I = 500$, but its performance quickly
deteriorates in smaller sample size because its $2$-norm
is quite large. When $I=100$, the best performer is $(8,7,8)$, which is
still monotone but less steep than $(20,20,20)$ as shown in
\Cref{fig:psi}.

When $F$ has a heavy tail such as $t_2$, it is
reasonable to expect that a redescending score function (such as
$(8,7,7)$ and $(8,6,7)$ as shown in \Cref{fig:psi}) yields a large
sensitivity value. This is confirmed by \Cref{tab:u2} in which
$(8,6,7)$ is the clear winner in all three sample sizes.

\renewcommand{\tabcolsep}{6pt}
\begin{table}[t]
  \centering
  \caption{Distributions of $\kappa^*_{0.05}$ for different U-statistics
    when $Y \sim \mathrm{N}(0.3, 1)$. Three sample sizes are
    considered: $I=100$, $I=500$ and $I = \infty$. In the first two
    sample sizes, we report the median and standard deviation in the
    normal approximation \eqref{eq:kappa-star-gamma}, as well as the
    median and standard deviation from $1000$ simulations which
    compute $\kappa^{*}$ using \eqref{eq:kappa-star-t}. The largest median in
    each column is bolded.}
  \label{tab:u1}
  \begin{tabular}{|l|rr|rr|r|}
    \hline
    \multirow{2}{*}{$(m, \underline{m}, \overline{m})$} & \multicolumn{2}{c|}{$I=100$} & \multicolumn{2}{c|}{$I = 500$} &
    \multirow{2}{*}{$I=\infty$} \\
    & approximation & simulation & approximation & simulation & \\
    \hline
    $(2,2,2)$  & $0.57~(0.0584)$ & $0.57~(0.056)$ & $0.623~(0.026)$ & $0.623~(0.0241)$ & $0.664~(0)$ \\
    $(8,8,8)$  & $0.581~(0.0943)$ & $0.585~(0.0956)$ & $0.677~(0.0423)$ & $0.681~(0.0421)$ & $0.748~(0)$ \\
    $(8,7,8)$  & ${\bf 0.587}~(0.0814)$ & ${\bf 0.595}~(0.0715)$ & $0.663~(0.0362)$ & $0.663~(0.0341)$ & $0.72~(0)$ \\
    $(8,6,8)$  & $0.582~(0.0708)$ & $0.587~(0.0688)$ & $0.648~(0.0314)$ & $0.646~(0.0291)$ & $0.698~(0)$ \\
    $(8,5,8)$  & $0.575~(0.0618)$ & $0.58~(0.0625)$ & $0.633~(0.0275)$ & $0.633~(0.0275)$ & $0.679~(0)$ \\
    $(20,20,20)$  & $0.528~(0.138)$ & $0.543~(0.131)$ & ${\bf
      0.681}~(0.0657)$ & ${\bf 0.683}~(0.0637)$ & ${\bf 0.791}~(0)$ \\
    $(20,18,20)$  & $0.576~(0.0962)$ & $0.579~(0.0987)$ & $0.678~(0.0433)$ & $0.677~(0.0436)$ & $0.753~(0)$ \\
    $(20,16,20)$  & $0.584~(0.0724)$ & $0.584~(0.0855)$ & $0.666~(0.0323)$ & $0.668~(0.0387)$ & $0.728~(0)$ \\
    $(8,7,7)$  & $0.568~(0.0761)$ & $0.569~(0.0729)$ & $0.638~(0.034)$ & $0.636~(0.0318)$ & $0.692~(0)$ \\
    $(8,6,7)$  & $0.559~(0.0666)$ & $0.554~(0.068)$ & $0.623~(0.0297)$ & $0.62~(0.0283)$ & $0.672~(0)$ \\
    \hline
  \end{tabular}
\end{table}

\begin{table}[t]
  \centering
  \caption{Distributions of $\kappa^*_{0.05}$ for different U-statistics
    when $Y \sim t_2 + 0.8$.
    Three sample sizes are
    considered: $I=100$, $I=500$ and $I = \infty$. In the first two
    sample sizes, we report the mean and standard deviation in the
    normal approximation \eqref{eq:kappa-star-gamma}, as well as the
    mean and standard deviation from $1000$ simulations which
    compute $\kappa^{*}$ using the first expression in \eqref{eq:kappa-star-t}. The largest mean/median in
    each column is bolded.
  }
  \label{tab:u2}

  \begin{tabular}{|l|rr|rr|r|}
    \hline
    \multirow{2}{*}{$(m, \underline{m}, \overline{m})$} & \multicolumn{2}{c|}{$I=100$} & \multicolumn{2}{c|}{$I = 500$} &
    \multirow{2}{*}{$I=\infty$} \\
    & approximation & simulation & approximation & simulation & \\
    \hline
    $(2,2,2)$  & $0.693~(0.0531)$ & $0.693~(0.0499)$ & $0.744~(0.023)$ & $0.744~(0.0239)$ & $0.781~(0)$ \\
    $(8,8,8)$  & $0.579~(0.0827)$ & $0.587~(0.105)$ & $0.676~(0.0371)$ & $0.678~(0.0434)$ & $0.747~(0)$ \\
    $(8,7,8)$  & $0.646~(0.0747)$ & $0.649~(0.0826)$ & $0.721~(0.0326)$ & $0.722~(0.0339)$ & $0.776~(0)$ \\
    $(8,6,8)$  & $0.681~(0.066)$ & $0.686~(0.0618)$ & $0.744~(0.0285)$ & $0.743~(0.0291)$ & $0.789~(0)$ \\
    $(8,5,8)$  & $0.698~(0.0615)$ & $0.697~(0.0526)$ & $0.754~(0.0265)$ & $0.753~(0.0243)$ & $0.794~(0)$ \\
    $(20,20,20)$  & $0.5~(0.147)$ & $0.5~(0.154)$ & $0.575~(0.0731)$ & $0.58~(0.073)$ & $0.691~(0)$ \\
    $(20,18,20)$  & $0.568~(0.118)$ & $0.578~(0.103)$ & $0.671~(0.053)$ & $0.674~(0.0461)$ & $0.746~(0)$ \\
    $(20,16,20)$  & $0.633~(0.0746)$ & $0.641~(0.0851)$ & $0.715~(0.0327)$ & $0.715~(0.0367)$ & $0.774~(0)$ \\
    $(8,7,7)$  & $0.689~(0.075)$ & $0.694~(0.0638)$ & $0.756~(0.0323)$ & $0.757~(0.0292)$ & $0.804~(0)$ \\
    $(8,6,7)$  & ${\bf 0.707}~(0.0588)$ & ${\bf 0.711}~(0.0643)$ &
    ${\bf 0.768}~(0.0252)$ & ${\bf 0.769}~(0.0257)$ & ${\bf 0.811}~(0)$ \\
    \hline
  \end{tabular}
\end{table}



In the supplementary file, we consider another class of test
statistics whose score functions are binary and obtained similar
conclusions to the U-statistics.

Based on the observations above, a reasonable strategy in practice is
to choose a statistic with large design sensitivity when the sample
size is large (e.g.\ $I \ge 500$). This can be done if prior knowledge
of the tail behavior is available, or a small planning sample can
be used to estimate $\mu_F$. In the latter case, a even better
strategy is to estimate the parameters $\mu_F$ and
$\sigma_F^2$ from the planning sample (for example by the Jackknife).
Then one can choose a statistic that maximizes the mean or some
quantile of the transformed sensitivity value $\kappa^{*}$ computed
from the theory-predicted distribution in \Cref{thm:kappa-star}.

\section{Selecting subpopulations}
\label{sec:select-subp}

In presence of effect modification (interaction between treatment
and covariates), \citet{hsu2013effect} discovered an interesting
phenomenon that the investigator might prefer to test on subgroups
with larger effects because they are less sensitive to hidden
bias. However, when the sample size is small, \citet{hsu2013effect}
found it more advantageous to use all the subgroups.

This phenomenon can be easily explained by the theoretical results in
\Cref{sec:main-result}, as the mean sensitivity value depends
on both the design sensitivity $\tilde{\Gamma} = \mu_F/(1-\mu_F)$ and
the sample size $I$. Suppose we have two subgroups whose tests statistics $T$ have
mean $\mu_{F1} > \mu_{F2}$
and the proportions of the two subgroups are $\pi_1$ and $(1 -
\pi_1)$, $0 < \pi_1 < 1$. When the sample size is sufficiently large,
\Cref{thm:kappa-star} implies that the transformed sensitivity value
obtained by using the first subgroup only converges to $\mu_{F1}$, and
the transformed sensitivity
value obtained by using both subgroups converges to $\mu_F = \pi_1
\mu_{F1} + (1 - \pi_1) \mu_{F2} < \mu_{F1}$. Therefore it is
preferable to use the first subgroup only. However, when the sample
size $I$ is small, using only $\pi_1$ proportion of the data is
less efficient and may produce smaller sensitivity value.

\begin{figure}[t]
  \centering
  \begin{subfigure}[t]{0.48\textwidth}
    \includegraphics[width = \textwidth]{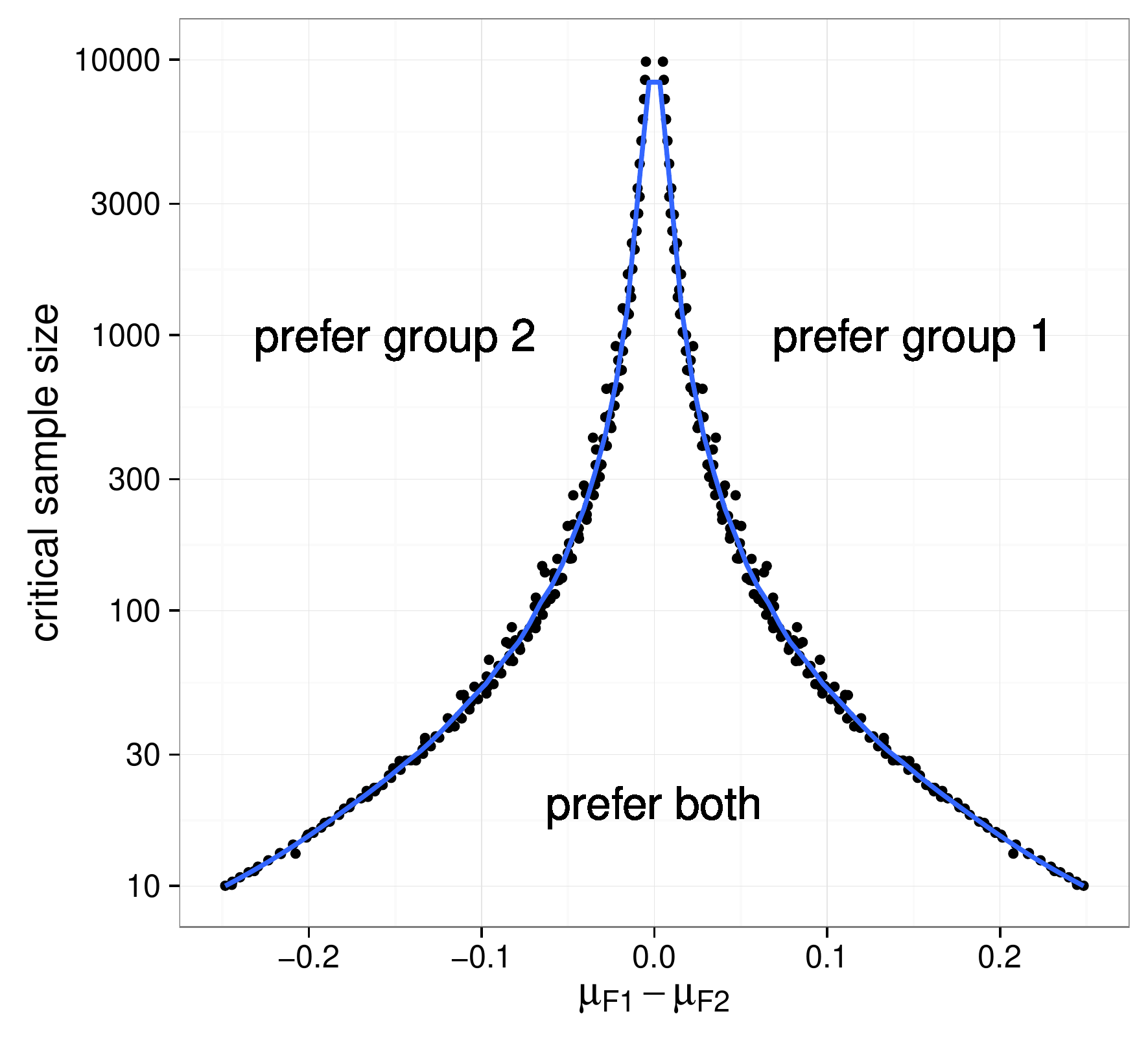}
    \caption{$\pi_1 = 0.5$.}
    \label{fig:lambda50}
  \end{subfigure} \quad
  \begin{subfigure}[t]{0.48\textwidth}
    \includegraphics[width = \textwidth]{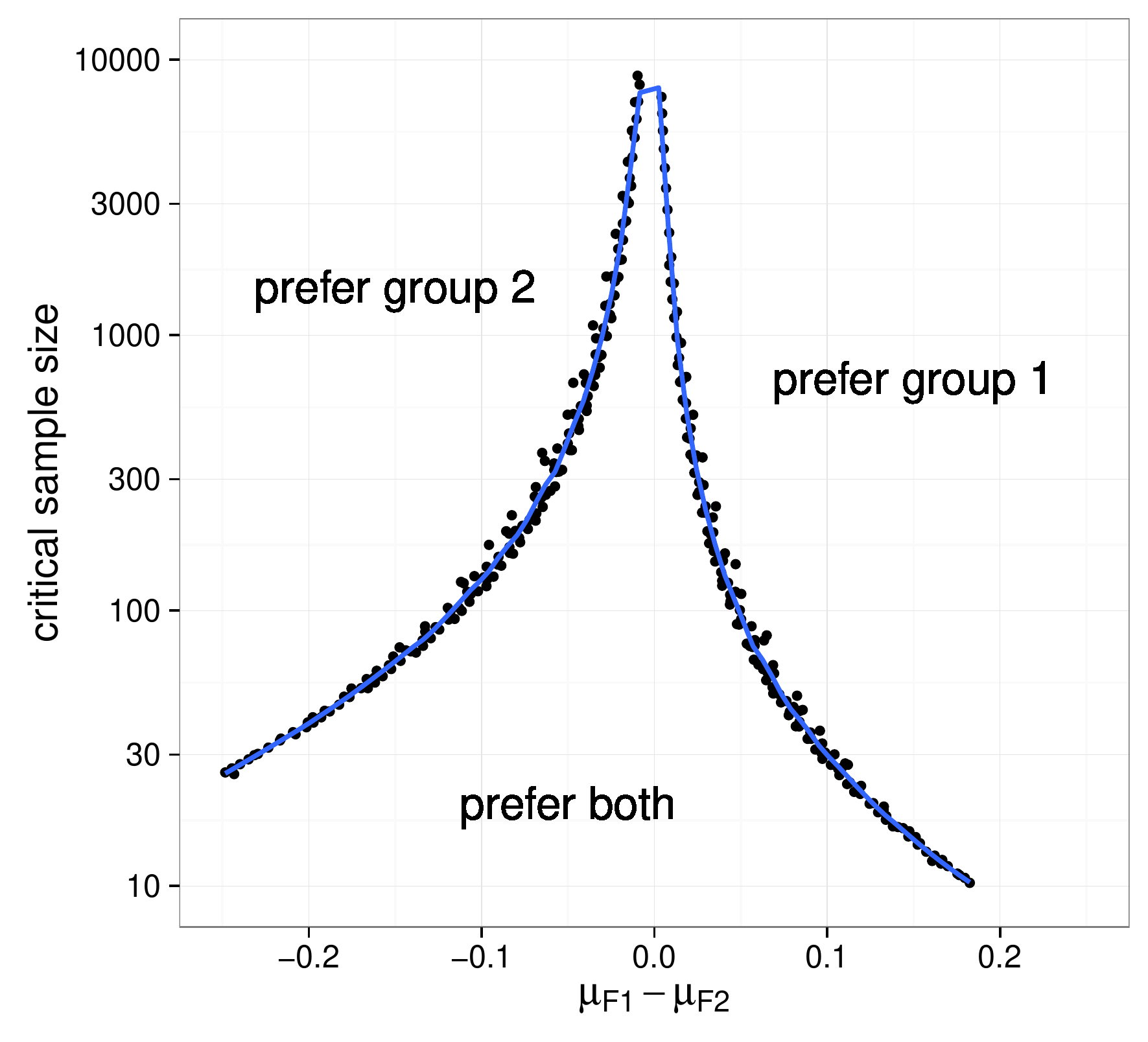}
    \caption{$\pi_1 = 0.75.$}
    \label{fig:lambda75}
  \end{subfigure}
  \caption{Tradeoff of sample size in selecting subpopulations in
    presence of effect modification. Each point corresponds to a
    combination of $\mu_{F1}$ and $\mu_{F2}$ between $0.5$ and $10/11$.}
  \label{fig:I-star}
\end{figure}

Next we compute the sample size threshold where the above transition
happens. Given
($\mu_{F1}$, $\mu_{F2}$, $\pi_1$), our goal is to determine the
critical sample size $I^{*}$ such that if $I > I^{*}$, using the
subgroup with larger effect gives larger transformed sensitivity value $\kappa^{*}$ on average,
and if $I < I^{*}$, using both groups gives larger transformed sensitivity value
$\kappa^{*}$ on average. Using the approximation \eqref{eq:kappa-star-gamma}, the
value $I^{*}$ can be determined by solving
\begin{equation} \label{eq:I-star}
  \begin{split}
    &\mu_F - \frac{(2 \mu_F - 1) \eta^{*} +
      \sqrt{4\eta^{*} \mu_F(1-\mu_F) +
        (\eta^{*})^2}}{2(1+\eta^{*})} \\
    =&  \mu_{F1} - \frac{(2 \mu_{F1} - 1)
      (\eta^{*} / \pi_1) +
      \sqrt{4(\eta^{*} / \pi_1) \mu_{F1}(1-\mu_{F1}) +
        (\eta^{*} / \pi_1)^2}}{2(1+\eta^{*}/\pi_1)},
  \end{split}
\end{equation}
where the root $\eta^{*} = \sigma_q^2 \bar{\Phi}^{-1}(\alpha)^2 /
I^{*}$. We numerically solve the equation for 50 equally spaced
$\mu_{F1}$ and $\mu_{F2}$ from $0.5$ to $10/11$ and $\pi_1 = 0.5$ and $0.75$. The critical
sample sizes $I^{*}$ are plotted in \Cref{fig:I-star} for the Wilcoxon's
test. For other statistics, $I^{*}$ can be obtained by multiplying a factor
that depends on their $\sigma_q^2$. Surprisingly, \Cref{fig:I-star}
shows $I^{*}$
primarily depend on the difference $\mu_{F1} - \mu_{F2}$. The curves in
\Cref{fig:I-star} define three regions where we would prefer the
first group, the second group, or both groups to minimize sensitivity
to unobserved bias. In practice, if effect modification is
expected to be substantial, one can estimate $\mu_{F1} - \mu_{F2}$
from a pilot sample and use \Cref{fig:I-star} to determine if just one
or both subgroups should be used.

\section{Selecting outcomes}
\label{sec:selecting-outcomes}

Lastly we consider observational studies with many outcomes of
interest. Our goal is to find the outcomes whose apparent
effects are least sensitive to unmeasured confounding. In these
problems, it is often helpful to reduce the number of outcomes, for
possibly two reasons:
\begin{enumerate}
\item In many problems, most of the outcomes have no or minuscule
  treatment effect. In this case, the $p$-value upper bound
  $\overline{p}_{\Gamma}$ is conservative if $\Gamma > 1$, and an unnecessary price
  of multiplicity is paid in multiple comparisons. Based on this observation,
  \citet{heller2009split} proposed a sample splitting procedure to
  screen out uninteresting outcomes and gain power; see also \citet{zhao2017cross}.
\item The observational study may simply be a preliminary study. In
  microarray studies, it is common to select some biomarkers (for example
  by a procedure controlling the false discovery rate) and see if they can
  be replicated in follow-up studies \citep{heller2014deciding}. 
\end{enumerate}

\begin{figure}[t]
  \centering
  \begin{tikzpicture}[node distance = 1.5cm, auto]
    \centering
    \node [block] (gender) {gender};
    \node [block, above right=of gender] (investigator) {investigator};
    \node [block, below right=of investigator, dashed] (batch) {batch, lab,
      microarray platform, etc.};
    \node [block, below right=of gender] (gene) {gene\\ expression};
    \path [line] (investigator) -- (gender);
    \path [line] (investigator) -- (batch);
    \path [line] (gender) -- (gene);
    \path [line] (batch) -- (gene);
  \end{tikzpicture}
  \caption{Illustration of unmeasured confounding in the gender
    study. The investigator analyze the samples in different
    batches, laboratories, or microarray platforms which may affect
    the gene expression. This introduces unmeasured confounding bias to
    the treatment effect.}
  \label{fig:batch-effect}
\end{figure}
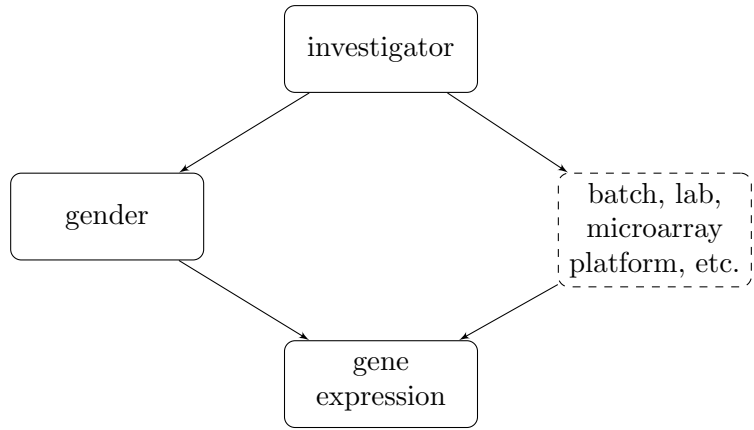

Sensitivity value is a natural way to screen the outcomes when we are
concerned about unmeasured confounding. Next we return to the genomics
example in \Cref{sec:introduction} and use the sensitivity value as an
exploratory tool.
In many microarray experiments, the target effects are
confounded by technical or non-biological
experimental variation when samples are processed in multiple
batches. \Cref{fig:batch-effect}
illustrates this source of unmeasured confounding. When some samples
are processed differently than others, for example, in different
laboratories or by different technicians, significant batch effects
may arise and confound the treatment effect we are interested in
\citep{leek2010tackling}. In the gender
example in \Cref{sec:introduction}, even after the observations are
matched by laboratory label and microarray platform, our results below
suggest that the study is still likely biased by other unmeasured
confounders.

Many statistical
methods have been proposed to adjust for the unmeasured confounding
\citep[e.g.][]{gagnon2012,sun2012}, but most of them need to
assume a linear model for the data (see \citet{wang2016confounder} for
an exposition). Sensitivity values provide a nonparametric and
computationally efficient way to
screen thousands of hypotheses. When unobserved confounding (e.g.\ batch
effect) is a major concern, we can compute a sensitivity value for each
hypothesis. Genes with extraordinarily large sensitivity values are more
likely to have genuine effects since their associations are less sensitive
to unobserved confounders.

\begin{figure}[htbp]
  \centering
  \includegraphics[width = \textwidth]{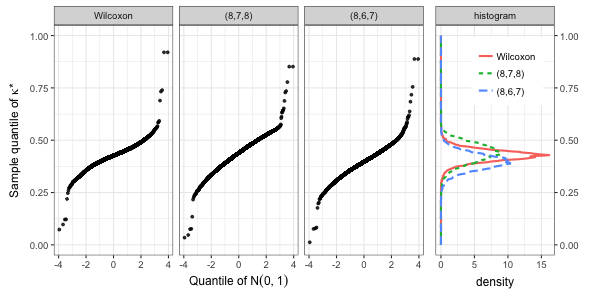}


  \caption{Using sensitivity values to screen genes in the microarray
    example (significance level $\alpha = 0.05$). The one-sided sensitivity
    values are computed with respect to the alternative
    hypothesis that male gene expressions are higher than female.}
  \label{fig:gene-screen}
\end{figure}

This new proposal is demonstrated in \Cref{fig:gene-screen} using the
gender example. We compute one-sided transformed sensitivity
values $\kappa^{*}_{0.05}$ with respect to the alternative that gene
expressions in males are higher. The left
three panels of \Cref{fig:gene-screen} show the
quantile-quantile (Q-Q) plot of the transformed sensitivity values
$\kappa^{*}$ versus the standard normal distribution for three
different test statistics. The rightmost panel shows the histogram
of the sensitivity values.

The empirical distribution of sensitivity values provides useful
information about genomics dataset, as it is usually safe to presume that most
genes have no or little genuine effects. In \Cref{thm:kappa-star}, the main
assumption is that the data are in the favorable situation, i.e.\
the random treatment assignment is satisfied after matching.
If that is true and the genes are independent, then by equation
\eqref{eq:kappa-star-gamma}, the empirical distribution of the sensitivity
values should be close to the normal distribution with mean $0.36$
for Wilcoxon, $0.31$ for $(8,7,8)$, or $0.33$ for $(8,6,7)$. The Q-Q plots in
\Cref{fig:gene-screen} show clear deviation from this theoretical
prediction: the empirical distribution have heavier tails and
the medians are different. This indicates unmeasured confounding
bias or very strong dependence between the genes (not very likely as
genetic dependence is usually local) or possibly both.

Among all the sensitivity values, a few of them are
clearly outliers. They correspond to the genes that are least
sensitive to unobserved bias and are more likely genuine effects. The
$9$ genes listed in \Cref{tab:gene} have two-sided
$\kappa^{*}_{0.05}$ greater than $0.6$ using Wilcoxon's
test ($5$ of them can be seen in \Cref{fig:gene-screen}). Among
them, $6$ are on the X/Y chromosome which are
more likely to be related to the gender as argued by
\citet{gagnon2012}. Using the method in
\citet{wang2016confounder} that estimates unmeasured confounders by
factor analysis and assumes genetic effects are sparse, all these $9$
genes have two-sided $p$-values less than $10^{-5}$.

\section{Discussion}
\label{sec:discussion}



Both sensitivity value and $p$-value are deterministic functions of
the data and indicate the level of confidence to reject the null
hypothesis. They are closely
related: both of them are increasing functions of the signed score statistic
$T$ if $T \ge 1/2$. In other words, they give essentially the same
ordering if we use them to screen outcomes as in
\Cref{sec:selecting-outcomes}.

However, $p$-value and sensitivity value are different transforms of
$T$ and should be used in
different study designs: $p$-value is only meaningful when there is no
unmeasured confounding, which is exactly what sensitivity value speaks
to. The distinction between sensitivity value and $p$-value is most clear if
we consider different test statistics. For example, Wilcoxon's test
has very good Pitman's efficiency for normal error \citep[Section
2.6]{hettmansperger1984statistical} but has poor efficiency in
sensitivity analysis as shown in \Cref{tab:u1} (the first row).
From the theoretical perspective, the distribution of the $p$-value is
commonly studied under local alternatives (e.g.\ location shift of the
order $1/\sqrt{I}$). For fixed alternative distributions, the $p$-value
in general converges to $0$ and does not carry much information. On the
contrary, the sensitivity value $\Gamma^{*}$ is not very meaningful under
local alternatives but behaves interestingly under fixed
alternatives as illustrated in this paper.



Throughout this paper, we have been working on pair-matched observational
studies and obtained some clean theoretical results. When there are
multiple controls for each treated observation, the theoretical
analysis becomes more difficult as there is no closed form solution
of the bounding variable $\overline{T}_{\Gamma}$, though it is
possible to find the asymptotic normal distribution of
$\overline{T}_{\Gamma}$ \citep{gastwirth2000asymptotic}. Of
course one can still compute the sensitivity value by binary-searching a
sensitivity analysis table, but it remains an open problem if there
exists a simple formula like \eqref{eq:kappa-star-t} for the
sensitivity value. A preliminary simulation study shows
that $\kappa^{*}$ is still asymptotically normal. We leave the
theoretical analysis for future research. More broadly, it would be
interesting to see if the concept of sensitivity value can extend to
other sensitivity analysis frameworks that do not assume homogeneous
treatment effect.

\section*{Acknowledgement}

The author thanks Paul Rosenbaum and Dylan Small for their
constructive suggestions.


\bibliographystyle{plainnat}
\bibliography{ref}

\appendix
\section{Proofs}
\label{sec:proofs}

\subsection{Proof of \Cref{thm:kappa-star}}
\label{sec:proof-crefthm:k-star}

Let $V_I = \sqrt{I} (T - \mu_F)$. By assumption, $V_I
\overset{d}{\to} \mathrm{N}(0, \sigma_F^2)$.
Since the limit of $\sigma_{q,I}^2$ exists and $\alpha$ is fixed, the
value $c$ converges to a constant $\sigma_q
\bar{\Phi}^{-1}(\alpha)$. Hence \eqref{eq:kappa-star-t} implies
\[
\begin{split}
  \kappa^{*} &= \frac{2IT - \sqrt{4c^2 I T (1-T)}}{2I} +
  o_p\Big(\frac{1}{\sqrt{I}}\Big) \\
  &= T - \frac{\sigma_q \bar{\Phi}^{-1}(\alpha) \sqrt{T(1-T)}}{\sqrt{I}} +
  o_p\Big(\frac{1}{\sqrt{I}}\Big) \\
  &= \mu_F + \frac{V_I}{\sqrt{I}} - \frac{\sigma_q \bar{\Phi}^{-1}(\alpha)
    \sqrt{\big(\mu_F + \frac{V_I}{\sqrt{I}}\big)\big(1-\mu_F -
      \frac{V_I}{\sqrt{I}}\big)}}{\sqrt{I}} +
  o_p\Big(\frac{1}{\sqrt{I}}\Big) \\
  &= \mu_F + \frac{V_I}{\sqrt{I}} - \frac{\sigma_q \bar{\Phi}^{-1}(\alpha)
    \sqrt{\mu_F(1 - \mu_F)}}{\sqrt{I}} +
  o_p\Big(\frac{1}{\sqrt{I}}\Big).
\end{split}
\]

Therefore
\[
\sqrt{I} (\kappa^{*} - \mu_F) = - \sigma_q \bar{\Phi}^{-1}(\alpha)
\sqrt{\mu_F(1 - \mu_F)} + V_I + o_p(1) \to \mathrm{N}(- \sigma_q \bar{\Phi}^{-1}(\alpha)
\sqrt{\mu_F(1 - \mu_F)}, \sigma_F^2).
\]

\subsection{Derivation of Equation \eqref{eq:kappa-star-gamma}}
\label{sec:proof-crefthm:k-star-1}

In \eqref{eq:kappa-star-gamma}, we assume $c^2 \approx \eta I$ instead
of a constant. By \eqref{eq:kappa-star-t}, we have
\[
\begin{split}
  \kappa^{*} &= \frac{2IT + \eta I - \sqrt{4\eta I^2 T (1-T) + \eta^2
      I^2}}{2(I + \eta I)} +
  o_p\Big(\frac{1}{\sqrt{I}}\Big) \\
  &= \frac{2\big(\mu_F + \frac{V_I}{\sqrt{I}}\big) + \eta - \sqrt{4\eta \big(\mu_F + \frac{V_I}{\sqrt{I}}\big) \big(1-\mu_F - \frac{V_I}{\sqrt{I}}\big) + \eta^2}}{2(1 + \eta)} +
  o_p\Big(\frac{1}{\sqrt{I}}\Big) \\
\end{split}
\]
Now use the Taylor expansion $\sqrt{a + x} = \sqrt{a} + x/(2\sqrt{a})
+ o(x)$, we get
\[
\kappa^{*} = \mu_F - \frac{(2\mu_F - 1) \eta}{2(1+\eta)} +
\frac{1}{\sqrt{I}} \frac{V_I}{1 + \eta} - \frac{\sqrt{4\eta
    \mu_F(1-\mu_F) + \eta^2} - \frac{4 \eta (2\mu_F - 1)}{2 \sqrt{4\eta
      \mu_F(1-\mu_F) + \eta^2}} \frac{V_I}{\sqrt{I}}}{2(1+\eta)} +
o_p\Big(\frac{1}{\sqrt{I}}\Big).
\]
Rearranging the terms, we get
\[
\begin{split}
  &\sqrt{I}\Big\{\kappa^{*} - \big[\mu_F - \frac{(2\mu_F - 1) \eta + \sqrt{4\eta
      \mu_F(1-\mu_F) + \eta^2}}{2(1+\eta)}\big]\Big\} \\
  =& \frac{V_I}{\sqrt{I}} \frac{1 + \frac{\eta (2\mu_F - 1)}{\sqrt{4\eta
        \mu_F(1-\mu_F) + \eta^2}}}{1+\eta} + o_p(1) \\
  \approx & \mathrm{N}\bigg(0,
  \frac{\sigma_F^2}{(1+\eta)^2} \Big( 1 +
  \frac{\eta(2\mu_F-1)}{\sqrt{4\eta \mu_F(1-\mu_F) + \eta^2}} \Big)^2
  \bigg).
\end{split}
\]


\section{Binary score functions}
\label{sec:quantile-tests}

To illustrate that larger design sensitivity does not always imply
larger sensitivity value in finite samples, consider the simple case
that $\psi$ is a binary function: for $0 \le \tau_u < \tau_l \le 1$,
\[
\psi(u) =
\begin{cases}
  1/(\tau_u-\tau_l) & \tau_l \le u \le \tau_u, \\
  0 & \mathrm{otherwise}. \\
\end{cases}
\]
This class of statistics generalize the sign test (corresponding to $\tau_l =
0$, $\tau_u = 1$) and were considered by \citet{noether1973some}. The
score function is already normalized such that $\|\psi\|_1 = 1$ and
note that $\|\psi\|_2 = 1 / (\tau_u - \tau_l)$.

\begin{figure}[t]
  \centering
  \begin{subfigure}[b]{0.49\textwidth}
  \includegraphics[width = \textwidth]{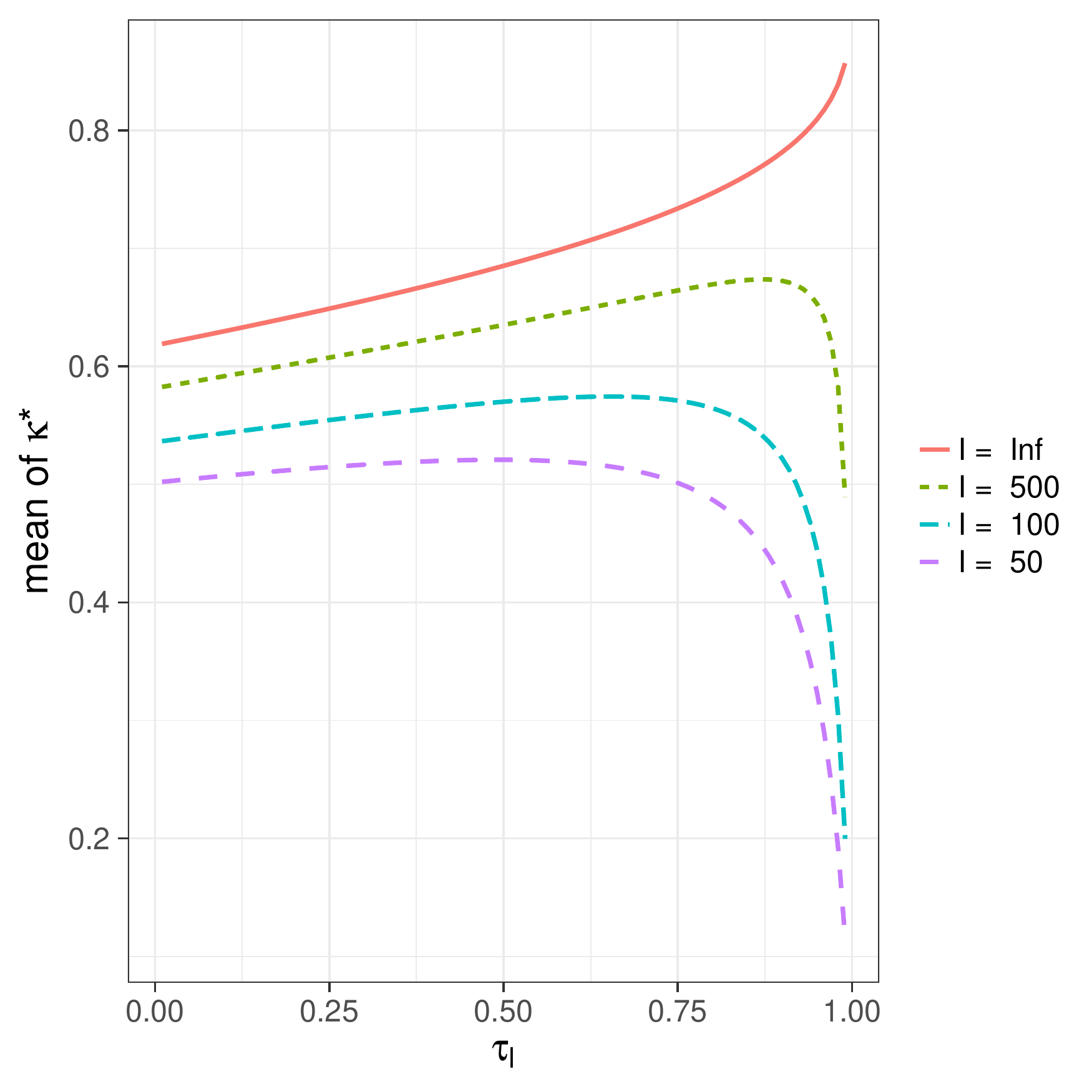}
  \caption{$Y \sim \mathrm{N}(0.3,1)$ and $\tau_u = 1$.}
  \label{fig:tau-normal}
  \end{subfigure}
  \begin{subfigure}[b]{0.49\textwidth}
  \includegraphics[width = \textwidth]{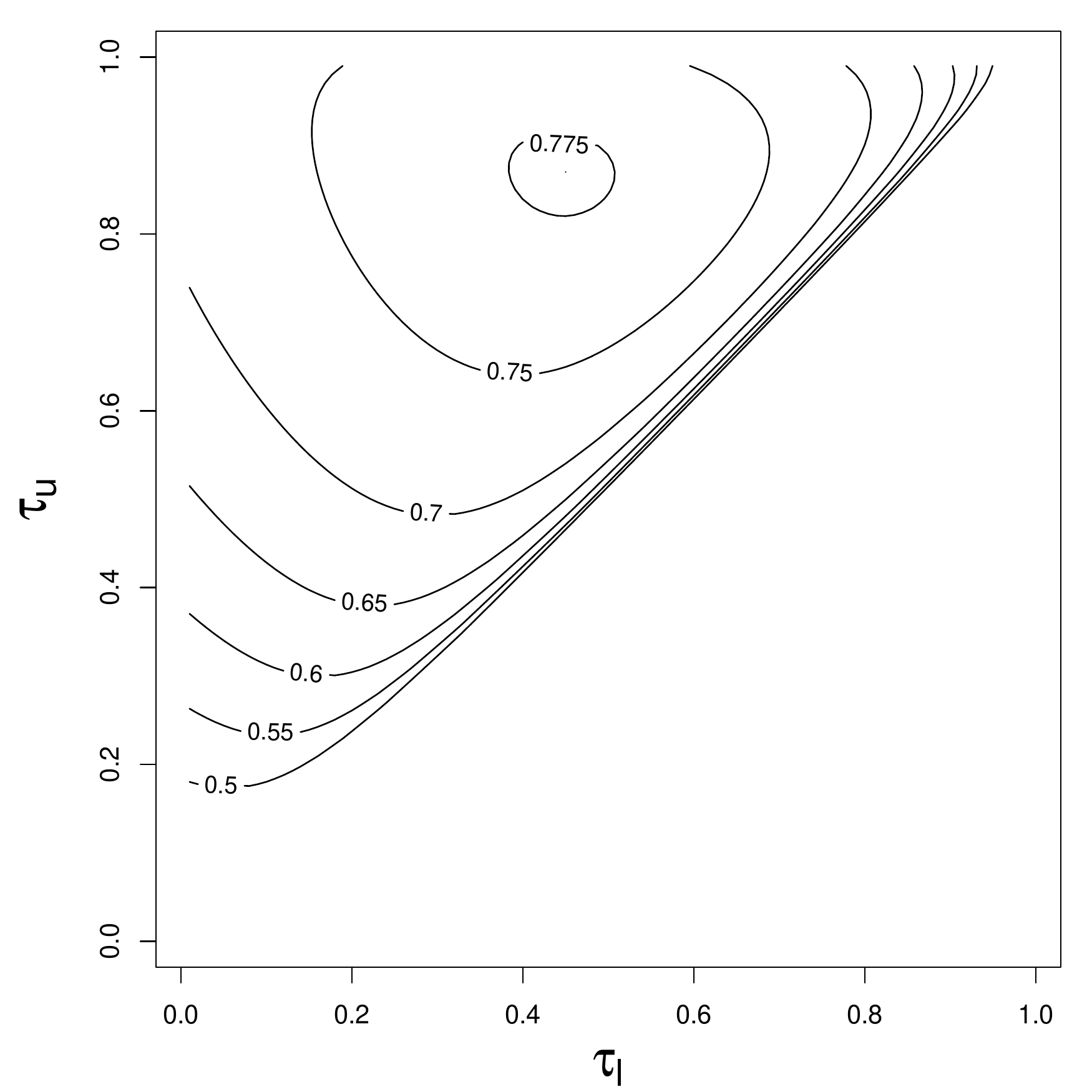}
  \caption{$Y \sim \mathrm{t}_2 + 0.8$ and $I = 500$.}
  \label{fig:tau-t2}
  \end{subfigure}
  \caption{The mean of sensitivity value versus different sample sizes
    and choices of $\tau_l$ and $\tau_u$.}
\end{figure}

As an illustration, suppose the data is generated by $Y_i
\overset{i.i.d.}{\sim} \mathrm{N}(0.3, 1)$. Since $g(u)$ is
increasing in this case, the optimal $\tau_u$ is $1$. We vary the
value $\tau_l$ over $(0,1)$ and plot in \Cref{fig:tau-normal} the
theoretical means of $\kappa^{*}$ (computed by
\eqref{eq:kappa-star-gamma}) for sample sizes $I = 50, 100, 500,
\infty$. When the sample size is finite, the mean of $\kappa^{*}$
starts to decrease as $\tau_l$ becomes close to $1$. This is expected
because $\|\psi\|_2 \to \infty$ as $\tau_l \to 1$.

When the tail of $F$ is heavy (such as $t_2$), it is sensible to
choose $\tau_u$ away from $1$. \Cref{fig:tau-t2} shows the contour
plot of the mean of $\kappa^{*}$ when $Y_i \overset{i.i.d.}{\sim} t_2
+ 0.8$, $I = 500$, and $(\tau_l, \tau_u)$ vary from $0$ to $1$. The
optimal binary score function is $\tau_l = 0.45$ and $\tau_u = 0.87$,
and the maximum mean of $\kappa^{*}$ is about $0.776$.


\end{document}